\documentclass[apjl,twocolumn]{emulateapj_mod}
\usepackage{epsfig}

\def\gtsima{$\; \buildrel > \over \sim \;$}
\def\ltsima{$\; \buildrel < \over \sim \;$}
\def\prosima{$\; \buildrel \propto \over \sim \;$}
\def\gsim{\lower.5ex\hbox{\gtsima}}
\def\lsim{\lower.5ex\hbox{\ltsima}}
\def\simgt{\lower.5ex\hbox{\gtsima}}
\def\simlt{\lower.5ex\hbox{\ltsima}}
\def\simpr{\lower.5ex\hbox{\prosima}}

\def\h1{$h^{-1}$}
\def\eeq{\end{equation}}
\def\beq{\begin{equation}}
\def\24mu{24\,$\mu{\rm m}$}
\def\70mu{70\,$\mu{\rm m}$}
\def\8mu{8\,$\mu{\rm m}$}
\catcode`\@=11
\def\gsim{\ifmmode{\mathrel{\mathpalette\@versim>}}
    \else{$\mathrel{\mathpalette\@versim>}$}\fi}
\def\lsim{\ifmmode{\mathrel{\mathpalette\@versim<}}
    \else{$\mathrel{\mathpalette\@versim<}$}\fi}
\def\@versim#1#2{\lower 2.9truept \vbox{\baselineskip 0pt \lineskip 
    0.5truept \ialign{$\m@th#1\hfil##\hfil$\crcr#2\crcr\sim\crcr}}}
\catcode`\@=12
\def\msun{\hbox{$M_\odot$}}

\submitted{Submitted to ApJ Letters on December 9, 2014; Accepted February  2, 2015}
\shorttitle{The Main Sequence of Star-Forming Galaxies}
\shortauthors{Renzini \& Peng}
\begin{document}

\title{An Objective Definition for the Main Sequence of Star-Forming Galaxies}

\author{
Alvio Renzini\altaffilmark{1},
Ying-jie Peng\altaffilmark{2}}

\altaffiltext{1}{INAF - Osservatorio
Astronomico di Padova, Vicolo dell'Osservatorio 5, I-35122 Padova,
Italy}  
\altaffiltext{2}{Cavendish Laboratory, University of Cambridge, 19 J. J. Thomson Avenue, Cambridge CB3 0HE, UK}

\email{alvio.renzini@oapd.inaf.it;  y.peng@mrao.cam.ac.uk}
 
\begin{abstract}
The Main Sequence (MS) of star-forming galaxies plays a fundamental role in driving galaxy evolution and in our efforts to understand it.
However, different studies find significant differences in the normalization, slope and shape of the MS. These discrepancies arise 
mainly from the different selection criteria adopted to isolate star-forming galaxies, that may include or exclude galaxies with specific  star formation rate (SFR)
substantially below the MS value. To obviate  this limitation of all current criteria, we propose an objective  definition of the MS that does not rely at all on a pre-selection of star-forming galaxies. Constructing the 3D SFR-Mass-Number plot, the MS is then defined as the ridge line of the star-forming peak, as illustrated with various figures. The advantages of  such definition are manifold. If generally adopted it will facilitate the inter-comparison of results from different groups using the same star formation rate (SFR) and stellar mass diagnostics, or to highlight the relative systematics  of different  diagnostics. All this  could help understanding MS galaxies as systems in a quasi-steady state equilibrium and would also provide a more objective criterion for identifying {\it quenching} galaxies.

\end{abstract}



\section{Introduction}
\label{intro}
The stellar mass and SFR of galaxies are 
fundamental quantities now being measured extensively, from low to the
highest redshifts at which galaxies have been discovered. For
star-forming (SF) galaxies the two quantities are tightly correlated with
each other, to the extent that, following \cite{Noeske07}, such a correlation has got the
designation of {\it Main Sequence} (MS) of star-forming galaxies. In a
series of seminal papers \citep{Noeske07,Daddi07,Elbaz07} it was recognized that such tight correlation
persists to at least redshift $\sim 2$ with nearly constant slope and
dispersion  compared to the correlation followed by SF galaxies in the
local Universe \citep{Brinchmann04}. Many subsequent studies have
followed,  confirming the existence of a MS, all the way to
at least $z\sim 4$ \citep{Pannella09, Peng10, Rodighiero10,
  Rodighiero11, Rodighiero14, Karim11, Popesso11, Popesso12, Wuyts11, Whitaker12, Whitaker14, Sargent12, 
  Kashino13, Bernhard14, Magnelli14}. Yet, slope, shape, dispersion and redshift
evolution of the SFR$-M*$ correlation can vary quite dramatically from
one study to another,  with the logarithmic slope of the relation
ranging  from $\sim 0.4$ up to $\sim 1$ (see e.g. the compilation in \cite{Speagle14}). In extreme cases, if galaxies
are collected in a SFR-limited fashion, no appreciable MS is recovered and the SFR
runs flat with stellar mass \citep{Erb06,Reddy06, Lee13}, a
result dominated by Malmquist bias  as at low stellar masses only galaxies with SFR above threshold are recovered \citep{ Rodighiero14}. 

\begin{figure*}[!]
\centering
\figurenum{1}
\epsscale{1}
\plotone{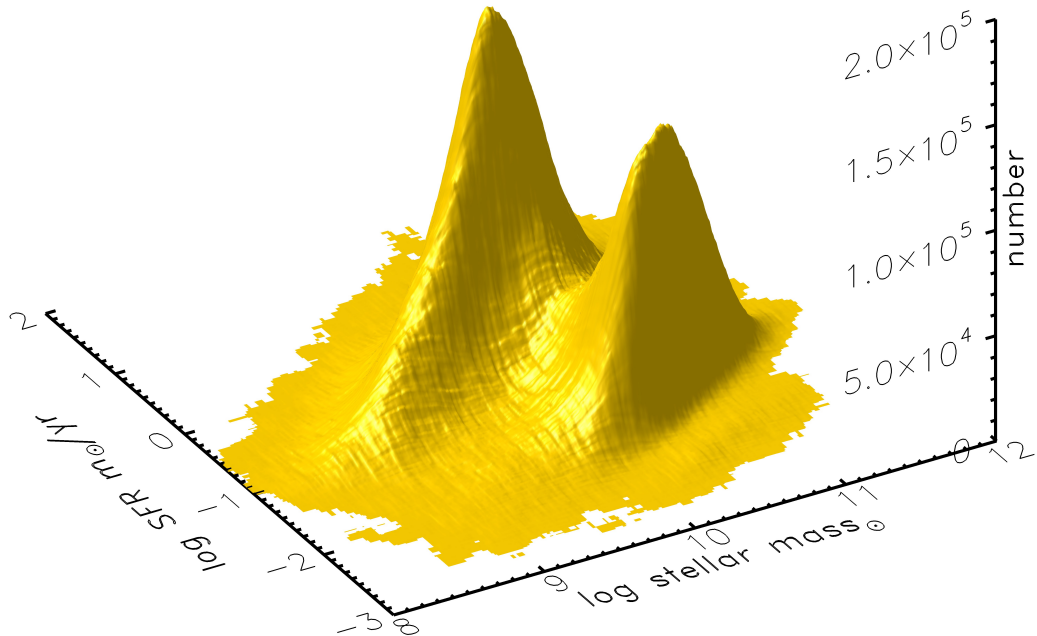}
\caption{The 3D SFR$-M*$ relation for local galaxies in the SDSS database
  and $0.02<z<0.085$. The third dimension is the
  number of 
galaxies in SFR-M bins. The drop towards lower masses is  partly
artificial, as no $V/V_{\rm max}$ correction has been applied.  This offers a
clearer vision 
of the 3D structure, with the two prominent peaks, one for
star-forming galaxies and one for the quenched ones.  Notice the sharp
ridge line of the SF peak, the extremely steep fall off in the number
of galaxies, either way of the ridge line, the divide,  which is then
been taken as the definition of the Main Sequence of star-forming
galaxies. On the North-West side of the divide one also encounter the
starburst outliers, at $z\sim 2$ include $\sim 2\%$ of galaxies and
contribute $\sim 10\%$ of the global SFR (Rodighiero er al. 2011). The SE side of
the divide is populated by a mixture of galaxies with lower SFR, with
some being just in a temporary excursion below  the MS band, while
others are definitely 
on their way across the saddle, towards the peak of quenched
galaxies. 
No $V/V_{\rm max}$ correction was applied in order to have a better visibility of the
two peaks. Data are from the SDSS database.}
\end{figure*}

One reason why the MS of one group may differ from that of
another stems primarily from how galaxies are selected for being
star forming. One may select them by a slightly mass-dependent color, picking {\it
blue cloud} galaxies (as in P10), or using the $BzK$ two-color selection \citep{Daddi04,Daddi07,Pannella09}, or the rest-frame $UVJ$ selection 
\citep{Williams09,Whitaker12, Whitaker14}, or setting
a minimum threshold for the specific SFR $\equiv$ SFR/$M*$  
\citep{Karim11}. Ultimately, all these criteria cut out galaxies with
low sSFR, but the threshold differ from one criterion to another, and
in a manner that may depend on mass, SFR, reddening  and redshift.
Criteria that tend to select as SF massive galaxies with low (but
detectable) SFR will return a flatter MS compared to criteria that
would not qualify as SF the same galaxies. Clearly, the use of
different criteria makes less comparable to each other the
corresponding results and may fuel sterile debates as to which of such
criteria should be preferable.

The importance of the MS comes from the fact that most stars in the Universe have
formed in galaxies lying around it within about a factor of $\sim$2 in SFR. The
mere existence of the MS, and its sharpness, indicate that there is order in nature, i.e., in the growth of galaxies, as opposed to mere stochasticity. 
It also implies that both the mass and SFR of high
redshift galaxies must increase with time quasi-exponentially
\citep{Renzini09, Maraston10}. The
absolute value of the sSFR sets the {\it clock} of galaxy evolution
(P10), setting the growth  rate of individual galaxies  and
controlling their lifetime  before they are quenched. The slope of the
MS  is directly connected to the (low-mass) slope of the mass function of
star-forming  galaxies, and if lower than unity would cause a runaway
steepening of the mass function if not contrasted 
by a tuned rate of merging \citep{Peng14}. Moreover, much interest
there is on the relation between the MS  and the specific
rate of mass increase of the dark matter halos hosting them, a
proxy of the rate of gas inflow  fueling the SF activity
\citep{Bouche10, Lilly13, PengM14}. Finally,  and perhaps most importantly, the SFR {\it distance} from the MS  can be used to identify galaxies 
that do not qualify as belonging to the MS, such as starburst outliers on one side and,  on the other side of the MS, those that have started the quenching process and, while still star forming, are in transition towards pure passive evolution.

For all these reasons, a more objective definition of
the MS of SF galaxies is in order: one which should not rely at all on
such a pre-selection. In this paper we accordingly propose a definition of the MS and suggest it
  should be generally adopted. 
Stellar masses and SFRs can be measured in many different ways,  e.g., for the SFR using  several different indicators through the whole electromagnetic spectrum, and the choice will depend on the available data and on the redshift of the sample. We emphasize that we are not proposing a universal way of measuring SFRs
(and masses), our goal is instead to objectively define the MS {\it once the measurements have been made}, whatever measurement procedures were used.  Still, such objective definition for the MS should also help quantify the relative systematics of different SFR and mass diagnostics.

\section{The 3D SFR-Mass-Number Plots}
We select galaxies from the Sloan Digital Sky Survey (SDSS) DR7
release \citep{Abazajian07} for lying at $0.02<z<0.085$ and having a
reliable spectroscopic redshift. Having excluded AGNs, this has provided a sample of $\sim
240,000$ galaxies for which SFRs  have been estimated from their
H$\alpha$ flux and their stellar masses from SED fits, following the same procedures as in  \cite{Brinchmann04} and P10,  with the exception that there SFRs were derived from  SDSS DR4 release. Galaxies were
weighted by the $V/V_{\rm max}$ method for masses below the
completeness limit. This dataset is ideal for our purposes, because of
its exquisite statistics, highly reliable redshifts and homogeneity in
SFR and mass estimates. However, the same procedure can be repeated at
higher redshifts as well, though other SFR diagnostics may have to be used.

Figure 1  shows the  3D SFR-Mass-Number plot, consisting
of the SFR$-M*$ relation where the third dimension gives the number of
galaxies in fixed-size (0.2$\times$0.2 dex) SFR$-M*$ bins. The two peaks correspond to
actively SF galaxies, on the {\it West} side, and quenched galaxies,
on the {\it East} side of the plot. Notice how sharp is the {\it divide} of the SF
peak. Besides error effects, the saddle between the two peaks must include a mixture of
galaxies, with some being in a temporary excursion to a low SFR
coming from the SF peak, others are in a temporary excursion from the
quenched peak, triggered by a minor gas-rich accretion event,  whereas others are truly quenching galaxies on their final journey from
the SF peak to the quenched galaxies depository. The number of quenched galaxies  increases continuously
from high to low redshifts, so across the valley the flow from the SF to the quenched peak must dominate.

In the 3D plot  of Figure 2  the third dimension
gives the product of the number of galaxies times their SFR, thus
illustrating with extreme clarity where in the SFR$-M*$ plane star
formation is concentrated.  Similarly, in  Figure 3 the third dimension now gives the product of the
number of galaxies times their mass, hence illustrating where 
the stellar mass resides.

In   Figure 2 the quenched galaxies virtually
disappear, as most of the star formation concentrates on the main
peak. The modest bump noticeable {\it South-East} of the main
SF peak is due to a low level of star formation still
present in a fraction of quenched galaxies, although just upper limits
to SFR might have been measured for many of them.  For this reason, the shape of the quenched peak  (especially on its low SFR side) should not be overinterpreted.  A comparison of the
two figures shows that the main peak has shifted from the star-forming
 to the quenched one, as indeed most stars in the local Universe are
 contained in quenched galaxies and bulges \citep{Baldry04}. 
 
A comparison of  these three figures reveals that there is a fair number of galaxies in the valley between the two peaks (Figure  1) that indeed contribute  to mass
(Figure 3) but not much to SFR (Figure 2).

\begin{figure*}
\centering
\figurenum{2}
\epsscale{0.9}
\plotone{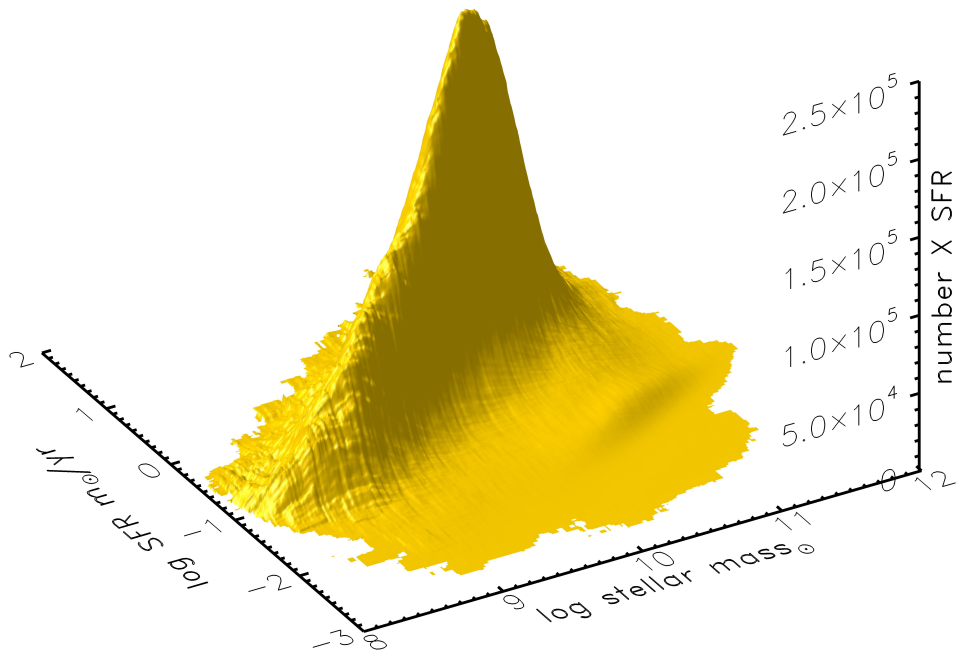}
\caption{Analog to Figure 1 where  the third dimension gives the
  product of the number $\times$ SFR, hence showing where most of the star
  formation 
takes place. The ridge line of the star-forming peak is the new
definition of the Main Sequence. The modest bump on the East side of
the main 
peak is due to quenched galaxies, some indeed still supporting a low level of star formation while data can give only upper limits for many of them. 
The $V_{\rm max}$ correction was applied for this plot.}
\end{figure*}

\begin{figure*}
\centering
\figurenum{3}
\epsscale{0.9}
\plotone{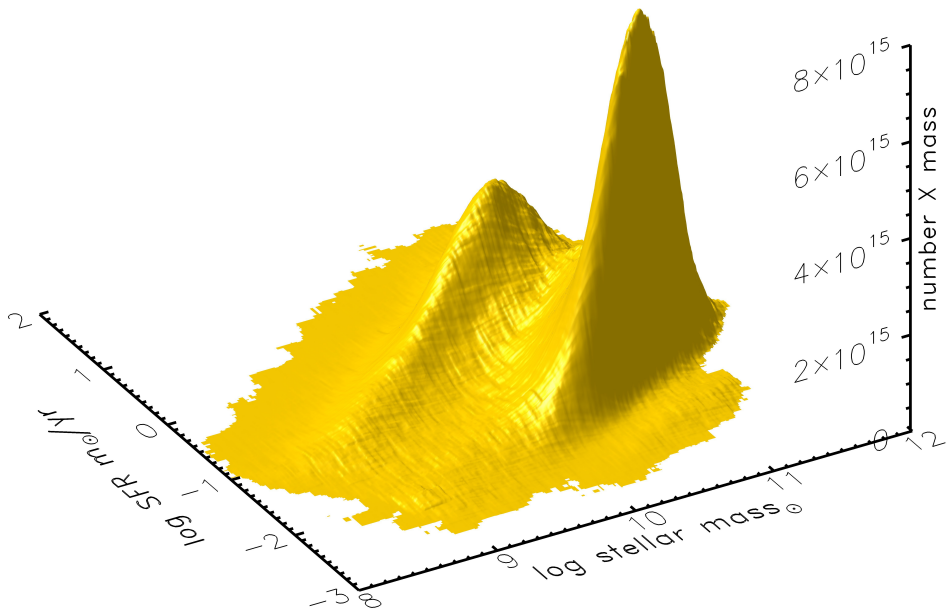}
\caption{The same of Figure 2, where  the third dimension now gives
  the product number $\times$  mass, hence showing where stellar mass is
  contained. The small SFR bump seen in Figure 2  has now exploded, as the majority of stars  in the local Universe reside in quenched galaxies. 
The $V_{\rm max}$ correction was applied for this plot.}
\end{figure*}

\section{An Objective  Main Sequence Definition}

As the importance of the MS stems from its dominant contribution to
the global star formation, it is worth rooting more deeply in this qualifying 
property its very definition. The sharpness of the SF peak 
in Figure 1 and Figure 2 offers indeed a quite 
natural definition of the MS as {\bf the ridge line of the SF peak}, either in the 3D number or in the 3D number$\times$SFR plots, as the two {\it divides} nearly overlap each other.
These 3D figures offer unquestionable evidence for the existence of the MS as well as of its more natural, cogent definition. This ridge line coincides with the {\it mode}  of the SFR distribution at fixed $M*$. The median or the average  SFR (at fixed $M*$)  might  be measured more accurately, but would depend on a pre-selection of star-forming galaxies. The ridge line, instead, does not require such pre-selection\footnote{The ridge/mode line was originally used by \cite{Brinchmann04} (see also 
\citealt{Salim07}) to fit the SF sequence in SDSS, but all subsequent studies adopted a pre-selection of SF galaxies to define the MS.}. Stacking e.g., infrared or radio data has been widely used to trace the MS (e.g., \citealt{Pannella09, Karim11, Rodighiero14}) but it requires pre-selecting SF galaxies, hence the result depends on the selection criterion.

Figure 4  shows the projection of the the 3D surface in Figure 1  over the SFR$-M*$ plane, along with contours for the number density of galaxies. Several notable features are emerging from this plot. The best straight-line fit to the ridge line is log(SFR) = $(0.76\pm 0.01){\rm log}(M*/\msun) - 7.64\pm 0.02$, with a slightly flatter slope than derived in P10 for $u-g$ color selected {\it blue} galaxies in the same catalog, which was $0.9\pm 0.01$. This  difference arises from the combination of two effects: the use of SDSS DR7 instead of the  DR4 release;  in addition,  the $u-g$  color cut included several galaxies with $M*\simeq 10^9\,\msun$ and with SFR definitely below the ridge line, that we may interpret as {\it quenching} galaxies.

Figure 4 shows that the ridge line is straight  linear up to the highest stellar masses in the sample, without a hint of  flattening  with increasing mass (see also \citealt{Brinchmann04}. A bending of the main sequence could be due to the growing fraction of the total mass being given by bulges already quenched, hence contributing mass but no star formation (e.g., \citealt{Whitaker12}). We don't see this effect in Figure 4 (up to $\sim 10^{11}\,\msun$), even though bulges should be maximally developed at $z\sim 0$.  Rather,  the increasing bulge fraction with galaxy mass may be responsible for the global deviation from $\sim 1$ of the slope of the SFR$-M*$ relation \citep{Abramson14}.

For galaxies away from the MS peak  Figure 4 shows that quenched galaxies populate  two  distinct peaks, one at high mass and one a low mass, which result from the {\it double-Schechter} shape of their mass function (\citealt{Baldry04}; P10). The high-mass peak is well separated from the MS, with very few galaxies with intermediate SFRs. The low-mass peak is instead connected in a continuous fashion to the MS, with many transition objects in between. This has a similar counterpart in the color-mass plot, where at low masses the distinction between blue and red galaxies gets blurred \citep{Taylor15}. In the frame of the P10 distinction between mass-quenched and environment quenched galaxies, the high-mass peak  can be  attributed to mass-quenching and the low-mass one to environment quenching.

The sharp dichotomy between the high mass, quenched galaxies and those populating the MS arises from two effects. The first, probably not the dominant one at this low redshift, is that the mass quenching process may be very fast, ensuring a rapid transition from the MS to the quenched galaxy repository. The second reason, that we believe is dominant at low redshifts, is that most massive galaxies were quenched a long time ago, so one expects very few to be in a transition phase now. Indeed, the stellar population properties of local massive early type galaxies (including ellipticals)  indicate that in most cases quenching took place some 10 Gyr ago (see e.g., \citealt{Renzini06}, for a review). This interpretation  is confirmed by the fact that virtually all massive ETGs are already in place at $z\sim 1$ \citep{cimatti06},  and therefore we  do not expect much mass quenching of massive galaxies  to take place now. 

\begin{figure*}[!]
\centering
\figurenum{4}
\epsscale{0.8}
\plotone{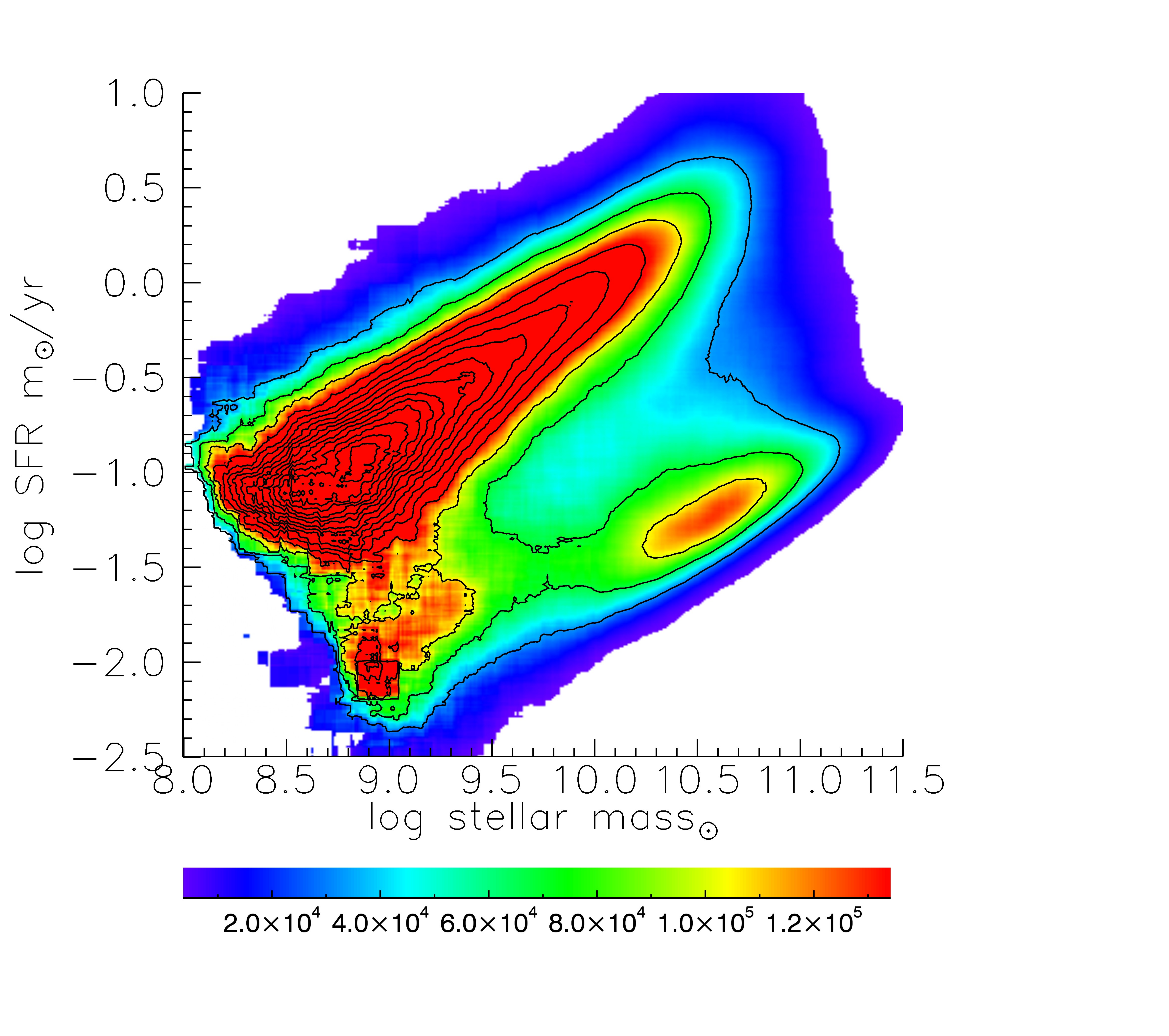}
\caption{The projection of the 3D surface shown in Figure 1  over the SFR$-M*$ plane. Level contours for the number density of galaxies are shown, with colors ranging from blue to red as a function of number density. The $V/V_{\rm max}$ correction was applied for this plot.}
\end{figure*}

The situation is just the opposite at low-masses. The environment quenching rate increases with decreasing redshift, following the growth of large scale structure 
overdensities, and therefore is maximum at redshift zero. Indeed, it is well know that the number density of quenched galaxies  increases by a factor of $\sim 2$ 
since $z\sim 1$ (e.g., \citealt{Bell04}), which is largely due to the increase in the number of low-mass quenched galaxies, whereas the number of massive quenched galaxies is stable in this redshift range. 

To illustrate our proposal, this paper is limited to use data relative to the local Universe, but it is  of great interests to follow the evolution of these 3D surfaces as a function of redshift, looking at how the {\it twin peaks} and  the galaxy population of the ({\it ``green''}) valley between them are changing with time.  Such surfaces can be used to trace the MS  divide at the various redshifts,
which should  be feasible even if current galaxy samples at high redshifts are not  as populous as those used here. This extension to higher redshifts is left to future works, but here we   can venture in making explicit a few expectations.  At redshift $\sim 2$ one expects to find  a very different pattern for quenched galaxies   compared to what  shown in Figure 4:  mass quenching must have started  at full steam (massive galaxies are growing very rapidly and very rapidly must be quenched). Instead, environment quenching has barely started, as overdensity contrasts are still small. One therefore expects the high-mass peak of quenched galaxies to be already in place and growing rapidly, whereas the low mass peak should  be barely noticeable. The flow of galaxies across  the {\it green valley}  should be high at high masses,  low at low masses.

These considerations may help understanding why in  $UVJ$-selected samples of SF galaxies \cite{Whitaker12} find a flattening of the SFR$-M*$ relation in high-mass galaxies at high redshifts, whereas \cite{Whitaker14} find a steepening of the relation for low-mass galaxies at low redshift. In the former case the flattening may be due to the inclusion of  massive galaxies already in their mass-quenching phase, whereas the steepening at low redshifts is likely due to the inclusion of low-mass galaxies in their environment-quenching phase  (such as those seen in Figure 4).


Finally, Figure 5 shows a section of the twin peaks at log$(M*)=10.5$, for both the number and the number$\times$SFR distributions. The shape of the distribution of the quenched peak is affected by the large number of galaxies for which just SFR upper limits. More interesting is instead the distribution of the star-forming peak. The $\sigma$ of its best fit Gaussian  is 0.3  dex, which comes from a combination of intrinsics spread and measurement errors \citep{Salmi12}.  We can notice two deviations from gaussianity in the wings of the distribution. On the low SFR side the excess w.r.t. the best fit Gaussian is  due to quenching galaxies, whereas at the opposite extreme the excess is likely due to starburst outliers from the MS. Notice also that there is a $\sim 0.2$ dex shift in the peak of the pure number distribution and the number$\times$SFR distribution, a shift that is  independent of mass: the ridge lines of the two surfaces run parallel to each other with a 0.2 dex difference and one can choose one or the other as the MS  divide. 

\begin{figure}[!]
\centering
\figurenum{5}
\epsscale{1.34}
\plotone{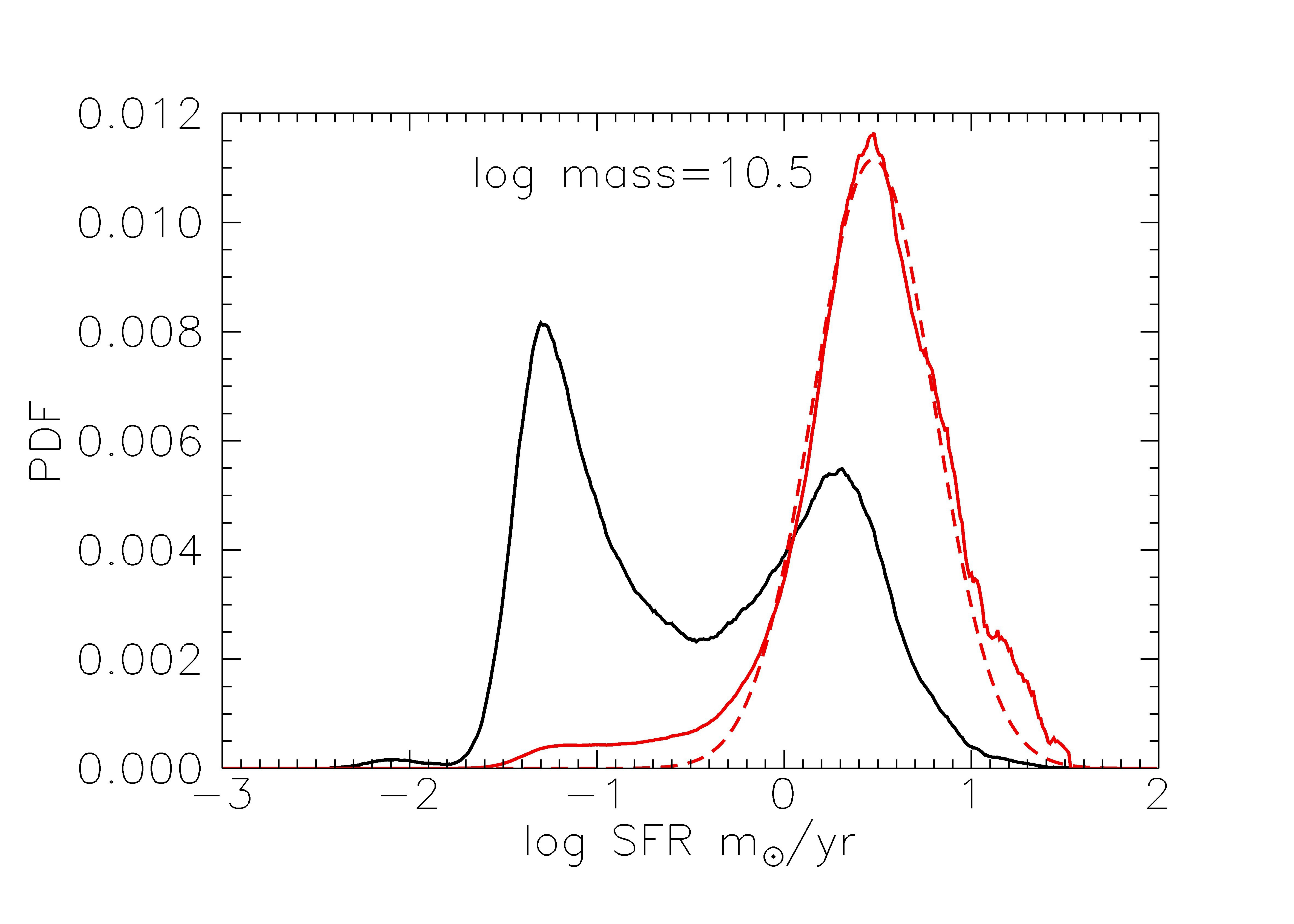}
\caption{A cut through the twin peaks at log$(M)=10.5$, in black showing the normalized probability distribution function (PDF) of the number of galaxies and in red the number$\times$SFR distribution (solid lines). The red dashed line shows the best fit Gaussian distribution.}
\end{figure}

\section{Conclusions}
We have proposed to objectively define the Main Sequence of star-forming galaxies as the ridge line in the 3D surface defined by the SFR-mass-number relation,
or nearly equivalently in the surface in which the third dimension is  the product number$\times$SFR. These surfaces can be constructed with very great 
statistical significance for the local sample of SDSS galaxies, and provide a vivid, cogent view of the reality of the MS as a major property of galaxy populations.
Such definition can be applied to samples galaxies at high redshifts as well, though with somewhat lesser statistical significance while different SFR indicators may have to be used in different redshift ranges. However, with the advent of near-IR multiobject spectrographs SFRs from H$\alpha$ can be derived up to $z\sim 2.5$, i.e., using the same indicator as used here for local galaxies (e.g., \citealt{Kashino13, Steidel14, Wisnioski14}). 

We emphasize that according to the new definition the MS of local, low$-z$ galaxies is indeed a straight line, with no sign of steepening at low masses or of flattening at high masses, features that may emerge when pre-selecting star-forming galaxies before constructing the   MS.  The logarithmic slope of the SFR$-M*$ relation is found to be 0.76 with the new definition of the MS, whereas it is 0.9 when using a $u-g$ color pre-selection as in P10, though part of the  difference comes from using   SDSS DR7 as opposed to DR4 data.

A projection of the SFR$-M*-$Number relation over the SFR$-M*$ plane reveals the existence of a number of low-mass galaxies with sub-MS SFRs, that we interpret as being undergoing environmental-quenching of their star formation, as expected in the phenomenological model of P10. At the opposite mass end, very few galaxies are now found in the course of their mass-quenching, as indeed  the model predicts that most of mass-quenching should have taken place at higher redshifts.

In summary, we propose a definition for the MS which does not require a pre-selection of SF galaxies, which should   facilitate the inter-comparison of results from different groups,  help understanding MS galaxies as systems in a quasi-steady state equilibrium and especially  provide a more objective criterion for identifying {\it quenching} galaxies.

\vskip - 0.9 cm
\section*{Acknowledgments}
    
We are grateful to Marcella Carollo, Natascha F\"orster Schreiber and Simon Lilly for useful discussions on these matters. AR acknowledges  support from a INAF-PRIN-2012 grant.


\label{lastpage}

\end{document}